\documentclass{amsart}

\usepackage{thumbpdf,lmodern}
\usepackage{amsmath,amssymb,amsfonts}
\usepackage{mathtools}
\usepackage{fancyvrb}
\usepackage{hyperref}
\usepackage{natbib}
\usepackage[utf8]{inputenc}   
\usepackage{url}
\usepackage{graphicx}
\usepackage{dsfont}
\usepackage{booktabs}

\newenvironment{Schunk}
  {\par\medskip}
  {\par\medskip}
\DefineVerbatimEnvironment{Sinput}{Verbatim}{fontshape=sl}
\DefineVerbatimEnvironment{Soutput}{Verbatim}{}
\DefineVerbatimEnvironment{Code}{Verbatim}{}
\DefineVerbatimEnvironment{CodeInput}{Verbatim}{fontshape=sl}
\DefineVerbatimEnvironment{CodeOutput}{Verbatim}{}

\newcommand{\code}[1]{\texttt{\detokenize{#1}}}
\let\proglang=\textsf
\newcommand{\pkg}[1]{{\fontseries{m}\fontseries{b}\selectfont #1}}

\begin{document}
\setkeys{Gin}{width=0.9\linewidth}
\title[\pkg{postcard} R package]{\pkg{postcard}: An R Package for Marginal Effect Estimation with or without Prognostic Score Adjustment} 
\author{Mathias Lerbech Jeppesen}

\author{Emilie Højbjerre-Frandsen*}


\date{\today}

\begin{abstract}
Covariate adjustment is a widely used technique in randomized clinical trials (RCTs) for improving the efficiency of treatment effect estimators. By adjusting for predictive baseline covariates, variance can be reduced, enhancing statistical precision and study power. \citet{RosenblumvanderLaan+2010} use the framework of generalized linear models (GLMs) in a plug-in analysis to show efficiency gains using covariate adjustment for marginal effect estimation.

Recently the use of prognostic scores as adjustment covariates has gained popularity. \citet{Schuler2020} introduce and validate the method for continuous endpoints using linear models. Building on this work \citet{HojbjerreFrandsen2025} extends the method proposed by \citet{Schuler2020} to be used in combination with the GLM plug-in procedure \citep{RosenblumvanderLaan+2010}. This method achieves semi-parametric efficiency under assumptions of additive treatment effects on the link scale. Additionally, \citet{HojbjerreFrandsen2025} provide a formula for power approximation which is valid even under model misspecification, enabling realistic sample size estimation.

This article introduces an R package, which implements the GLM plug-in method with or without PrOgnoSTic CovARiate aDjustment, \pkg{postcard}. The package has two core features: (1) estimating marginal effects and the variance hereof (with or without prognostic adjustment) and (2) approximating statistical power. Functionalities also include integration of the Discrete Super Learner for constructing prognostic scores and simulation capabilities for exploring the methods in practice. Through examples and simulations, we demonstrate \pkg{postcard} as a practical toolkit for statisticians.
\end{abstract}

\maketitle

\section{Introduction}

Covariate adjustment in the analysis of randomized clinical trials (RCTs) is a widely used technique for improving the efficiency of treatment effect estimators. By adjusting for predictive baseline covariates we reduce the variance of estimators, thereby enhancing statistical efficiency and increasing study power. For outcomes modeled with linear regression, incorporating baseline covariates into the statistical analysis has been shown to substantially reduce variance \citep{Tsiatis2008, RosenblumvanderLaan+2010}. For binary outcomes, \citet{MoorevanderLaan+2009} demonstrated precision gains through covariate adjustment with logistic regression. These findings were generalized by \citet{RosenblumvanderLaan+2010} using the framework of generalized linear models (GLMs) \citep{Nelder1972} in a plug-in analysis, making covariate adjustment for marginal effect estimation accessible for a broader array of data types.

However, including a large number of covariates can increase the risk of overfitting, potentially inflating the type I error rate. Furthermore, post hoc selection of adjustment covariates may introduce biases, violating the principle of pre-specification that regulatory agencies such as the U.S. Food and Drug Administration (FDA) and European Medicines Agency (EMA) advocate \citep{FDA_cov, EMA_cov}. To address these issues and to incorporate knowledge from previous trials, researchers have turned to prognostic scores; a single covariate constructed from historical data that provide a way to harness predictive information from previous trials without introducing bias into the analysis \citep{Holzhauer2022, Schuler2020}.

Prognostic scores, often derived using data from historical controls, combine multiple baseline covariates into a single prognostic covariate. By adjusting for this combined measure during the statistical analysis, precision of the statistical analysis can be enhanced. Linear adjustment for prognostic scores has recently received significant attention due to its theoretical and practical advantages. For continuous outcomes, \citet{Schuler2020} demonstrated that linear adjustment with a prognostic score achieves asymptotic efficiency under the assumption of constant treatment effect. Moreover, the method adheres to regulatory guidance on pre-specification of covariates and avoids bias introduced by post hoc adjustments. \citet{HojbjerreFrandsen2024} gives guidelines on practical implementations and considerations for this method. \citet{HojbjerreFrandsen2024} suggest to model the prognostic scores using a Discrete Super Learner \citep{superlearner}. 

Building on the foundational work of \citet{Schuler2020}, the method has been extended to the GLM plug-in procedure for estimating marginal effect measures by \citet{HojbjerreFrandsen2025}. By incorporating prognostic scores into the GLM plug-in procedures for marginal effects estimation, we achieve semi-parametric efficiency under assumptions of additive treatment effects on the link scale. These findings broaden the applicability of prognostic score adjustment and address gaps in clinical trial methodology, particularly for discrete and non-continous outcome types. Additionally, \citet{HojbjerreFrandsen2025} provide a formula for conservatively estimating the asymptotic variance, facilitating power calculations that reflect these efficiency gains even when the GLM model is misspecified.

\begin{figure}[h]
\centering
\includegraphics[width=0.95\textwidth]{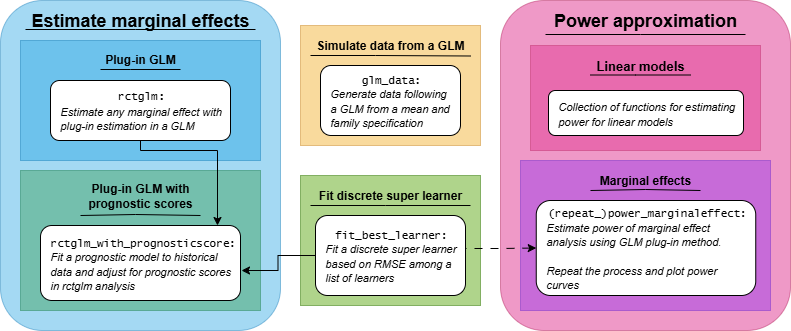}
\caption{Overview of most important functionalities in the \pkg{postcard} package. Abbreviations: Generalized linear model (GLM) and root mean squared error (RMSE).}
\label{fig:pkg-overview}
\end{figure}

To bring these advances into practice, we have developed the R package \pkg{postcard} that implements the GLM plug-in procedure introduced by \citet{RosenblumvanderLaan+2010} for marginal effect estimation with or without prognostic score adjustment. The R software package provides a practical toolkit for statisticians. There are two main features of the R package: (1) estimating marginal effects and their associated variances, and (2) approximating statistical power for prospective sample size estimation. Additionally, the package includes functionalities for data simulation and fitting a Discrete Super Learner (see Figure~\ref{fig:pkg-overview}). 

In Section~\ref{sec:stat_methods}, we present the statistical foundation underpinning these two primary functionalities. This section provides a concise summary of the theoretical framework proposed by \citet{HojbjerreFrandsen2025}, which serves as the methodological basis for the package. 

In Section~\ref{sec:est_marg_package}, we turn our attention to the package functionalities related to estimating marginal effects (depicted within the first box of Figure~\ref{fig:pkg-overview}). Section~\ref{sec:marg_without} explores marginal effect estimation without adjustment for a prognostic score, providing a default application of the GLM plug-in method. Section~\ref{sec:marg_with} then introduces functionalities using prognostic score adjustment, which involves a dependency on the Discrete Super Learner used for prognostic model building. The package allows users to fit a Discrete Super Learner to flexibly construct highly predictive prognostic scores from baseline covariates, as indicated by the arrow linking the Discrete Super Learner to marginal effect estimation in Figure~\ref{fig:pkg-overview}.

In Section~\ref{sec:pros_power_package}, we focus on the package functionalities for power approximation, with particular emphasis on the novel procedure proposed by \citet{HojbjerreFrandsen2025} for power approximation in the presence of model misspecification. While the package supports power calculations under the assumption of a correctly specified linear model, this research highlights the application to GLM plug-in analyses with or without prognostic score adjustment. As with marginal effect estimation, this power approximation functionality relies on the Discrete Super Learner for scenarios involving prognostic adjustment, as reflected by the dashed arrow between the Discrete Super Learner and the power approximation node in Figure~\ref{fig:pkg-overview}.

To illustrate the functionalities of the package, we simulate data throughout the article using a function provided within the package itself (see the orange box in Figure~\ref{fig:pkg-overview}). The simulated data follow a GLM model. While this simulation functionality is a secondary aspect of the package, it serves as a practical tool for showcasing the package’s core features.

postcard is a specialized framework for RCTs with the objective to estimate marginal effect estimands within a GLM based modeling setup. In contrast to the broader tlverse ecosystem that utilises the tmle3 R package. In tmle3 the focus is on implementing the TMLE framework itself rather than on implementing a specific TML estimator and a specific causal model like an RCT framework. Our framework in postcard is deliberately smaller and more problem specific. Concretely, we restrict attention to an RCT structural setup, which implies that the estimator is constructed so that it already solves the relevant estimating equation, and therefore does not require an additional updating/targeting step to achieve the desired inference. Instead we have included the use of historical data by modelling and including estimated prognostic scores. While tlverse relies on the sl3 R package to implement super learning as part of its general TMLE nuisance learning strategy, postcard restricts the trial outcome model component to GLMs and uses the super learner only for prognostic model fitting on historical data. The resulting prognostic scores are then included as an additional covariate to improve estimation of the outcome model in the RCT analysis. Thus, historical information enters through prognostic augmentation of the outcome model, rather than through a fully general super learner based pipeline.

Also, postcard provides a capability that is particularly useful in the design stage of a trial: a procedure for power/sample-size estimation targeted to marginal effect estimation under GLMs. By connecting the marginal effect estimand, its IF based variance, and resulting power calculations, postcard supports trial planning in a way that is more directly matched to estimating marginal effect rather than traditional power approaches. In comparison, packages such as pwrss are largely oriented toward more classical testing and power formulations under narrower statistical assumptions, and they do not provide the same workflow.

There are other packages within the causal estimation realm. The R package marginaleffects \cite{JSSv111i09} also work with marginal effects, but it targets different need compared to postcard. The marginaleffects R package is primarily about computing and interpreting marginal effects from fitted models, whereas postcard provides an RCT focused framework for marginal effect inference (based on the IF variance), including historical data in the form of prognostic scores and power/sample-size estimation for marginal effect estimands under GLMs. The CovSel R package \cite{JSSv068i01} is aimed at data driven covariate selection to improve causal effect estimation, whereas postcard does not focus on covariate selection, since the statistical analysis in many pharmaceutical settings needs to be pre-specified before analyzing the data. 

\section{Statistical methods}\label{sec:stat_methods}

\subsection{Marginal effect definition}\label{sec:def-margeffect}
We work in the setting of a two-armed trial with $n$ participants in total, where each observational unit is denoted as $O=\left (W,A,Y\right)$ and are drawn independently from the unknown distribution $\mathcal{P}$. Here, $Y$ is the endpoint variable, which can be continuous, binary, or ordinal, and $W$ represents a vector of baseline covariates. Participants are randomly allocated to either the treatment ($A=1$) or control group ($A=0$), with treatment probabilities $P\left(A = a \right) = \pi_a$ where $0<\pi_a<1$ for $a\in \{0,1\}$. We denote the sizes of the two treatment groups as $n_1$ and $n_0$ for the treatment and control groups, respectively.

Each participant has the two potential outcomes: the outcome under treatment, $Y(1)$, and the outcome under control, $Y(0)$ \citep{Petersen+vdLaan2014, Imbens2004, Sekhon2008}. Then the population mean outcome under treatment $a$ can be denoted as $\Psi_a=\mathbb{E}[Y(a)]$. Standard identification arguments show $\Psi_a=\mathbb{E}\mathbb{E}[Y(a)|W] = \mathbb{E}\mathbb{E}[Y|A = a, W]$. We can then use the notation $\mu(a, W)= \mathbb{E}[Y|A=a, W]$ and $\mu_a(W) =\mathbb{E}[Y(a)|W]$ interchangeably.

Our focus is on marginal effects, which are defined as causal effects in the form
\begin{align}\label{eq:def_r}
\Psi = r\left(\Psi_1, \Psi_0\right),
\end{align}
where $r$ is continuously differentiable in $(\Psi_1, \Psi_0)$ and 
\begin{align}\label{eq:asump_r}
r_0'(\Psi_0,  \Psi_1) \coloneqq \frac{\partial r}{\partial \Psi_0}(\Psi_0,  \Psi_1) \leq 0 \quad \wedge \quad r_1'(\Psi_0,  \Psi_1) \coloneqq \frac{\partial r}{\partial \Psi_1}(\Psi_0,  \Psi_1) \geq 0.
\end{align}

Commonly used effects are the average treatment effect (ATE), rate ratio (RR) and odds ratio (OR), and information about these marginal effects can be seen in Table~\ref{tab:marginal_effects}. 

\begin{table}[h]
\centering
\begin{tabular}{p{0.2\textwidth} p{0.2\textwidth} p{0.5\textwidth}}
\toprule
Marginal effect & $r\left(\Psi_1, \Psi_0\right)$ & Interpretation \\
\midrule
ATE & $\Psi_1 - \Psi_0$ & Difference in counterfactual group means. Often used for continuous data modelled with a linear regression in the GLM plug-in procedure.\\
RR & $\frac{\Psi_1}{\Psi_0}$ & Ratio of counterfactual group means. Often used for count data modelled with Poisson or Negative binomial regression in the GLM plug-in procedure. \\
OR & $\frac{\Psi_1 / (1- \Psi_1)}{\Psi_0 / (1-\Psi_0)}$ & Odds ratio for counterfactual group means. Often used for binary data modelled with logistic regression in the GLM plug-in procedure. \\
\bottomrule
\end{tabular}
\caption{Examples of marginal effects}
\label{tab:marginal_effects}
\end{table}

\subsection{Estimating marginal effects with GLMs}\label{sec:marg-eff}

A GLM estimates the conditional mean $\mu(A, W)$ using a link function $g$ to relate the mean to a linear combination of predictors, $\beta_0 1_n + x \beta_x$, where $(1\,\, x)$ is a row in the $n\times (1+q)$ design matrix $\mathbb{X}$ using a column of ones, $1_n$, alongside $A$, $W$ and interaction effects. The design matrix must include $1_n$ and the treatment variable $A$. Collapsibility problems implies that the coefficients of the GLM are usually not interpretable as direct estimators of the marginal effects \citep{Huitfeldt2019, Rhian2021}. Instead, \citet{RosenblumvanderLaan+2010} suggest to use a simple and general plug-in method using GLMs in the setting of an RCT. \citet{RosenblumvanderLaan+2010} demonstrate that the GLM-based plug-in estimator qualifies as a regular and asymptotically linear estimator (RAL) regardless of model misspecification, only assuming the independence of $W$ and $A$. This is because it is as a targeted maximum likelihood estimator (TMLE) completed in zero steps. Being RAL means that the estimator remains consistent and has a asymptotically normal sampling distribution, with its variance determined by the variance of the influence function (IF). \citet{RosenblumvanderLaan+2010} assume standard regularity conditions \citep{R+vdL2009} to ensure convergence of parameter estimates for the GLMs. They also assume that the GLM uses the canonical link functions and stem from either Normal, Binomial, Poisson, Gamma, or Inverse Normal distributions. \citet{HojbjerreFrandsen2025} extend this to also include the Negative Binomial distribution. The procedure can be summarized in the following steps: 
\begin{enumerate}
\item Estimate the maximum likelihood estimate (MLE) $\hat{\beta}$ and then estimate the conditional mean functions as $\hat{\mu}(a, w) = g^{-1}\left(\hat{\beta}_0 + x\hat{\beta}_x\right)$.
\item Extract the counterfactual predictions from the GLM, first assuming everyone in the sample was actually treated ($a=1$), and then assuming the opposite ($a=0$), i.e.
\begin{align*}
\hat{\Psi}_a = \frac{1}{n}\sum_{i=1}^n \hat{\mu}(a, W_i),  \,\,\,\,\,\,\,\,\,\, a \in \{0, 1\}. 
\end{align*}
\item\label{itm:estimand} Get the marginal effect estimate on the original scale by plugging in 
\begin{align*}
\hat{\Psi} = r\left(\hat\Psi_1, \hat{\Psi}_0\right).
\end{align*}

\item\label{itm:asympvar} The sampling variance of the IF gives a consistent estimate of the asymptotic variance. Using the delta method \citep{vdV98} gives, 
\begin{align*}
\hat{v}_\infty^2 = \frac{1}{n}\sum_{i=1}^n \left( r_0'(\hat\Psi_1, \hat\Psi_0)\cdot\hat\phi_{0}(A_i, W_i, Y_i)  + r_1'(\hat\Psi_1, \hat\Psi_0)\cdot\hat\phi_{1}(A_i, W_i, Y_i) \right)^2,
\end{align*}
where $\hat\phi_{a}(A_i, W_i, Y_i) = \dfrac{I(A_i = a)}{\pi_a}(Y-\hat\mu(a, W_i))+(\hat\mu(a, W_i)-\hat\Psi_a)$ and $r'_a$ is defined in Equation~\ref{eq:asump_r}. One can construct confidence intervals (CI) and perform hypothesis testing using this estimate of the IF. 
\begin{enumerate}
\item[(4a)] A cross-validation approach can be utilized to account for the modeling process. The dataset is divided into training and validation sets, adhering to the randomization ratio. The GLM is applied to the training data, and the IF is computed for the validation data. The sample variance can then be derived from this cross-validated IF estimate.
\end{enumerate}
\end{enumerate}

\subsubsection{Leveraging a prognostic model}\label{sec:prog-model}

We now assume that we have access to data from $\tilde{n}$ historical control group participants. \citet{Schuler2020} applied the idea of a prognostic score by using it as an additional adjustment covariate in a linear model, effectively decreasing the CIs while minimizing the risk of overfitting. A tutorial on the use of linear adjustment for a prognostic score can be found in \citep{HojbjerreFrandsen2024}. The prognostic score is defined as 
\begin{align}
\rho\left(W\right) \coloneqq \mathbb{E}[Y \,|\, W,A=0, D=0],
\end{align}
where $D$ is an indicator for being in the new trial. The prognostic score is estimated by training a machine learning model such as a Discrete Super Learner on the historical data,  \citep{superlearner}. We denote the estimate as $\hat{\rho}(W)$. \citet{HojbjerreFrandsen2025} propose to use a prognostic score to increase the efficiency of the marginal effect estimate. This is done by adding $g(\hat{\rho}(W))$ as an additional covariate in the design matrix in the GLM plug-in procedure. \citet{HojbjerreFrandsen2025} show that including this as an additional covariate results in local efficiency of the marginal effect estimator when the true treatment effect is additive on the link scale, i.e. $g\left(\mu(1, W)\right) = \zeta + g\left(\mu(0, W)\right)$. Using simulated data \citet{HojbjerreFrandsen2025} show that one can expect an efficiency increase even when the treatment effect is not additive on the link scale. 

\subsection{Prospective power estimation}\label{seq:pro-power}
\citet{HojbjerreFrandsen2025} describe a procedure for conducting a conservative prospective power estimation which is valid even under model misspecification. The procedure can be summarized as: 
\begin{enumerate}
\item Fix $\pi_a$ and target effect size $\Psi$.
\item Gather historical data that accurately reflects the design, population, and data type of the new study. 
\item Use the historical data to estimate  $r_a'$, $\sigma_a^2 = \mathbb{V}\mathrm{ar}\left[Y(a)\right]$ and
$$\kappa_a^2 = \mathbb{E}\left[(Y(a)-\mu^*(a, W))^2\right],$$
where $\mu^*(a, W))$ is the large sample GLM fit. For details see \citep{HojbjerreFrandsen2025}. 
\item Find the variance bound by plugging the estimates into 
\begin{align*}
v_\infty^{\uparrow 2} = r_0'^{\, 2}\sigma_0^2+ r_1'^{\, 2}\sigma_1^2+ \pi_0\pi_1\left(\frac{|r_0'|\kappa_0}{\pi_0} + \frac{|r_1'|\kappa_1}{\pi_1} \right)^2.
\end{align*}
\item\label{itm:crit_val} Let $\Delta$ be the marginal treatment effect in case of no treatment difference and $\alpha$ be the significance level. 
Let $\mathcal{H}_0: \hat{\Psi} \sim F_0 = \mathcal{N}(\Delta ,v_\infty^{\uparrow 2} / n)$ and $\mathcal{H}_1: \hat{\Psi} \sim F_1 = \mathcal{N}(\Psi,v_\infty^{\uparrow 2} / n)$. Increase $n$ until 
\begin{align*}
1-\gamma = 1- F_1\left(F_0^{-1}(1-\alpha/2)\right)    
\end{align*}
exceeds the desired power level.
\end{enumerate}

\section[Estimating marginal effects using postcard]{Estimating marginal effects using \pkg{postcard}}\label{sec:est_marg_package}

The \pkg{postcard} package implements the methods outlined in \citep{HojbjerreFrandsen2025} and briefly described in the previous section to estimate marginal effects and estimate study power, potentially utilizing historical data to increase accuracy of estimation. Figure~\ref{fig:pkg-overview} provides an overview of the package functionalities as described in the introduction. Here we will go through the first of the two main functionalities of the package, namely estimating marginal effects using the GLM plug-in procedure. 

\pkg{postcard} is available for download on CRAN and can be downloaded using the code:

\begin{Schunk}
\begin{Sinput}
R> install.packages("postcard")
\end{Sinput}
\end{Schunk}

The development version is available on \href{https://github.com/NovoNordisk-OpenSource/postcard}{GitHub}. The development version of the package can be installed using the code:

\begin{Schunk}
\begin{Sinput}
R> pak::pak("NovoNordisk-OpenSource/postcard")
\end{Sinput}
\end{Schunk}

Note that the version used for generation of this article is from the latest tagged development release (\href{https://github.com/NovoNordisk-OpenSource/postcard/releases/tag/v1.1.0.9000-paper}{v1.1.0.9000-paper}).

To be able to showcase the functionalities of the package, a function to simulate data from a GLM, \code{glm_data()}, is exported in the package. The user can specify an expression alongside variables and a family of the response to then simulate a variable from a GLM with linear predictor given by the expression provided.

\subsection[Estimating marginal effects with rctglm()]{Estimating marginal effects with \code{rctglm()}}\label{sec:marg_without}

The function \code{rctglm()} implements the procedure described in section \ref{sec:marg-eff} and returns an S3 object of class \code{rctglm} that most importantly contains a data set with the estimated marginal effect in item~(\ref{itm:estimand}) in section~\ref{sec:marg-eff} alongside an estimated standard error (SE) derived from the equation in item~(\ref{itm:asympvar}). This is accessible via S3 methods \code{estimand()} or \code{est()}. A breakdown of other elements of an \code{rctglm} object is available in the \code{Value} section of the documentation.

While documentation is available in the package, we briefly explain the role of the arguments here.

\subsubsection{Default behavior}

As a default, \code{rctglm()} estimates ATE as the marginal effect in a linear model. \code{rctglm()} has four arguments with no default value, and these need to be specified by the user. First, the argument \code{formula} describes the GLM model to be fitted, exactly like in the \code{glm()} function from the \pkg{stats} package included in the R distribution \citep{rcore2025}. Next is \code{exposure_indicator} which is the (name of) the binary group indicator in data equivalent to $A$ as described in section~\ref{sec:def-margeffect}. \code{exposure_prob} is the probability of being in the treatment group, i.e. $\pi_1$. Last argument with no default value is \code{data}, which is a specific data set that at least contains the columns specified in \code{formula}.

We simulate data from a Gaussian as well as a Poisson distribution to be able to showcase both default and non-default behavior of \code{rctglm()}.

\begin{Schunk}
\begin{Sinput}
R> n <- 1000
R> b0 <- 1
R> b1 <- 3
R> b2 <- 2
R> # Simulate data with a non-linear effect
R> dat_gaus <- glm_data(
+    Y ~ b0+b1*log(X)+b2*A,
+    X = runif(n, min = 1, max = 50),
+    A = rbinom(n, 1, prob = 1/2)
+  )
R> dat_pois <- glm_data(
+    Y ~ b0+b1*log(X)+b2*A,
+    X = runif(n, min = 1, max = 50),
+    A = rbinom(n, 1, 1/2),
+    family = poisson(link = "log")
+  )
\end{Sinput}
\end{Schunk}

The \code{rctglm()} function estimates any specified estimand using plug-in estimation and estimates the variance using the variance of the IF. If we only specify arguments with no default values, the function estimates the ATE in a linear model. An example showcasing this using the above data is seen below:

\begin{Schunk}
\begin{Sinput}
R> ate <- rctglm(formula = Y ~ A * X,
+                exposure_indicator = A,
+                exposure_prob = 1/2,
+                data = dat_gaus)
\end{Sinput}
\end{Schunk}

The output is an object of class \code{rctglm} that contains the estimated marginal effect, the SE, and information on the underlying GLM fit. The object prints as:

\begin{Schunk}
\begin{Sinput}
R> ate
\end{Sinput}
\end{Schunk}

\begin{Schunk}
\begin{Soutput}
Object of class rctglm

Call: rctglm(formula = Y ~ A * X, exposure_indicator = A,
exposure_prob = 1/2,
data = dat_gaus)

Counterfactual control mean (psi_0=E[Y|X, A=0]) estimate: 9.993
Counterfactual active mean (psi_1=E[Y|X, A=1]) estimate: 11.97
Estimand function r: psi1 - psi0
Estimand (r(psi_1, psi_0)) estimate (SE): 1.974 (0.08698)
\end{Soutput}
\end{Schunk}

The estimated marginal effect (in this case ATE) can be accessed using the \code{estimand()} (or shorthand \code{est()}) function, which returns a data frame with the estimand and its SE.

\begin{Schunk}
\begin{Sinput}
R> est(ate)
\end{Sinput}
\begin{Soutput}
  Estimate Std. Error
1 1.974324 0.08698096
\end{Soutput}
\end{Schunk}

Note that the estimate $1.97$ is close to the true treatment effect of $2$.

It is also possible to retrieve coefficient information from the underlying GLM fit with the \code{coef} function.

\begin{Schunk}
\begin{Sinput}
R> coef(ate)
\end{Sinput}
\begin{Soutput}
(Intercept)           A           X         A:X 
 5.90940625  2.11271778  0.16075185 -0.00544781 
\end{Soutput}
\end{Schunk}

A \code{summary} method gives a coefficient-style table for the estimand with its standard error, a Wald test against the null value of the estimand (0 for the ATE) and a confidence interval. The confidence level can be changed with the \code{level} argument.

\begin{Schunk}
\begin{Sinput}
R> summary(ate)
\end{Sinput}
\end{Schunk}

\begin{Schunk}
\begin{Soutput}
Summary of rctglm fit

Call: rctglm(formula = Y ~ A * X, exposure_indicator = A,
exposure_prob = 1/2,
data = dat_gaus)

Counterfactual control mean (psi_0=E[Y|X, A=0]): 9.993
Counterfactual active mean (psi_1=E[Y|X, A=1]): 11.97

Estimand function r: psi1 - psi0
Null value (r(psi, psi)): 0

Estimate Std. Error z value Pr(>|z|) 2.5 % 97.5 %
Estimand 1.97432 0.08698 22.7 4.651e-114 1.804 2.14

(Wald-type 95% CI; standard error computed from the full-model
influence-function variance.)
Access the underlying GLM summary via `<summary_object>$glm_summary`.
\end{Soutput}
\end{Schunk}

A \code{plot} method also exists for visualising the estimate and its confidence interval, with a dashed line marking the null value.

\begin{Schunk}
\begin{Sinput}
R> plot(ate)
\end{Sinput}
\end{Schunk}
\includegraphics{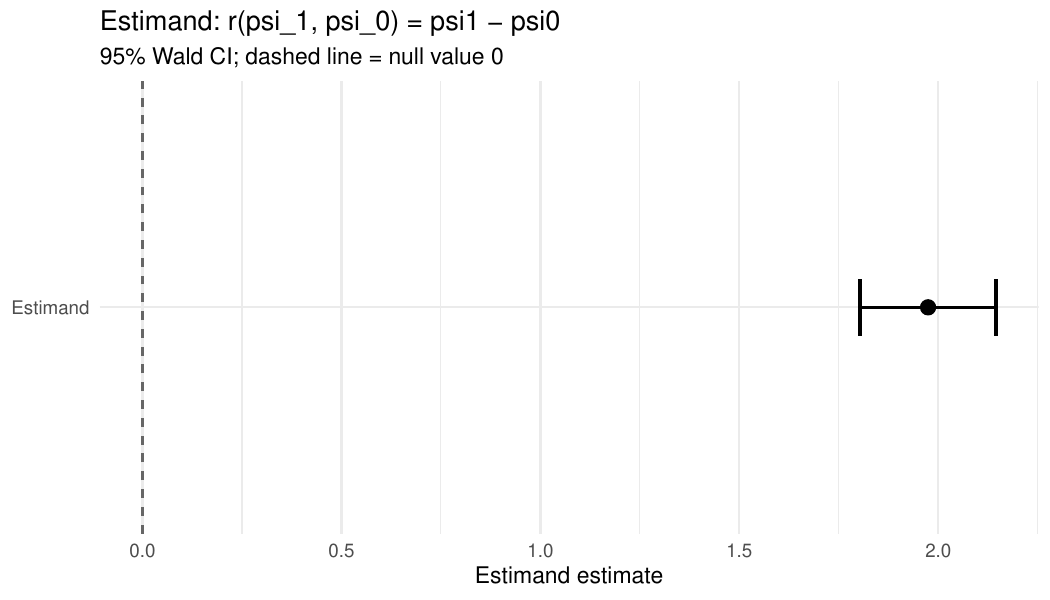}

\subsubsection{Changing default behavior}

\code{rctglm()} has additional arguments to alter the default behavior of the function. This includes a \code{family} argument specifying the error distribution and link function to use in the GLM, equivalent to the same argument in the \code{glm()} function. Ellipses \code{...} are available to pass other arguments on to the \code{glm()} function. An \code{estimand_fun} argument determines the function $r$ in Equation~\eqref{eq:def_r}, meaning it determines what type of marginal effect estimand is estimated by the procedure. The default is \code{"ate"}, which produces a difference in means estimand $\Psi_1-\Psi_0$. Another acceptable character input is \code{"rate_ratio"}, which produces the estimand $\frac{\Psi_1}{\Psi_0}$. Code below shows this:

\begin{Schunk}
\begin{Sinput}
R> rr <- rctglm(
+    formula = Y ~ A + X,
+    exposure_indicator = A,
+    exposure_prob = 1/2,
+    data = dat_pois,
+    family = "poisson",
+    estimand_fun = "rate_ratio")
\end{Sinput}
\end{Schunk}

\begin{Schunk}
\begin{Sinput}
R> rr
\end{Sinput}
\end{Schunk}

\begin{Schunk}
\begin{Soutput}
Object of class rctglm

Call: rctglm(formula = Y ~ A + X, exposure_indicator = A,
exposure_prob = 1/2,
data = dat_pois, family = "poisson", estimand_fun = "rate_ratio")

Counterfactual control mean (psi_0=E[Y|X, A=0]) estimate: 94856
Counterfactual active mean (psi_1=E[Y|X, A=1]) estimate: 702052
Estimand function r: psi1/psi0
Estimand (r(psi_1, psi_0)) estimate (SE): 7.401 (0.07446)
\end{Soutput}
\end{Schunk}

Though these two frequently used estimands are easily available as character strings, the normal input type for the \code{estimand_fun} argument is a function with arguments \code{Psi0} and \code{Psi1}, allowing the user full flexibility in the type of marginal effect estimand to use. Below we show an example of a \code{function} specification of the estimand.

\begin{Schunk}
\begin{Sinput}
R> estimand_fun <- function(psi1, psi0) {
+    psi1 / sqrt(psi0) * 2 - 1
+  }
R> nse <- rctglm(
+    formula = Y ~ A * X,
+    exposure_indicator = A,
+    exposure_prob = 1/2,
+    data = dat_pois,
+    family = poisson(),
+    estimand_fun = estimand_fun)
\end{Sinput}
\end{Schunk}

\begin{Schunk}
\begin{Sinput}
R> nse
\end{Sinput}
\end{Schunk}

\begin{Schunk}
\begin{Soutput}
Object of class rctglm

Call: rctglm(formula = Y ~ A * X, exposure_indicator = A,
exposure_prob = 1/2,
data = dat_pois, family = poisson(), estimand_fun = estimand_fun)

Counterfactual control mean (psi_0=E[Y|X, A=0]) estimate: 94857
Counterfactual active mean (psi_1=E[Y|X, A=1]) estimate: 702052
Estimand function r: {
psi1/sqrt(psi0) * 2 - 1
}
Estimand (r(psi_1, psi_0)) estimate (SE): 4558 (83.48)
\end{Soutput}
\end{Schunk}

Arguments \code{estimand_fun_deriv0} and \code{estimand_fun_deriv1} are the functional derivatives of the function $r$ with respect to $\Psi_0$ and $\Psi_1$, respectively. As a default these are \code{NULL}, which means these are derived from the specified \code{estimand_fun} using symbolic differentiation available in the \pkg{Deriv} package \citep{Deriv2024}. The user can choose to specify them manually as a \code{function} with arguments \code{Psi0} and \code{Psi1}. If verbose is $\geq 1$, information on the derivation of these derivatives is printed to the console. It is also possible to retrieve information on the estimand function as well as the derivatives by inspecting the \code{estimand_funs} element of the \code{rctglm} class object as seen below:

\begin{Schunk}
\begin{Sinput}
R> nse$estimand_funs
\end{Sinput}
\begin{Soutput}
$f
function(psi1, psi0) {
  psi1 / sqrt(psi0) * 2 - 1
}

$d0
function (psi1, psi0) 
-(psi1/(psi0 * sqrt(psi0)))

$d1
function (psi1, psi0) 
2/sqrt(psi0)
\end{Soutput}
\end{Schunk}

A logical argument \code{cv_variance} is available to use cross validation to estimate the variance and inherently the SE as described in bullet (4a) in section~\ref{sec:marg-eff}. The default is \code{FALSE}. The \code{cv_variance_folds} is the number of folds to use for the cross validation if \code{cv_variance} is \code{TRUE}, which as a default is set to \code{10}. Below we see the result of re-running the first instance of \code{rctglm} but with \code{cv_variance = TRUE}.

\begin{Schunk}
\begin{Sinput}
R> cvvar <- rctglm(formula = Y ~ A * X,
+         exposure_indicator = A,
+         exposure_prob = 1/2,
+         data = dat_gaus,
+         cv_variance = TRUE,
+         cv_variance_folds = 5)
\end{Sinput}
\end{Schunk}

\begin{Schunk}
\begin{Sinput}
R> cvvar
\end{Sinput}
\end{Schunk}

\begin{Schunk}
\begin{Soutput}
Object of class rctglm

Call: rctglm(formula = Y ~ A * X, exposure_indicator = A,
exposure_prob = 1/2,
data = dat_gaus, cv_variance = TRUE, cv_variance_folds = 5)

Counterfactual control mean (psi_0=E[Y|X, A=0]) estimate: 9.993
Counterfactual active mean (psi_1=E[Y|X, A=1]) estimate: 11.97
Estimand function r: psi1 - psi0
Estimand (r(psi_1, psi_0)) estimate (SE): 1.974 (0.08729)
\end{Soutput}
\end{Schunk}

Lastly, a \code{numeric} argument \code{verbose} determines the level of verbosity, i.e. how much information about the procedure is printed in the console to the user, with \code{0} being no verbosity. The default value is set as \code{2} via an option in the package, which for example prints information about the symbolic differentiation of the estimand function. See more information in the help page by running \code{?postcard::options}.

\subsection{Leveraging a prognostic model}\label{sec:marg_with}

As described in section~\ref{sec:prog-model}, a marginal effect can be estimated where we adjust for a prognostic score created by the use of a prognostic prediction model fitted to historical data. This procedure is implemented in the function \code{rctglm_with_prognosticscore()}.

\code{rctglm_with_prognosticscore()} fits a prognostic model to historical data and adjusts for the prognostic scores in a call to \code{rctglm} to estimate the marginal effect and its SE. Hence, this function has all the same arguments as \code{rctglm()} but with additional arguments needed to fit the prognostic model. These include the historical data on which to fit the prognostic model, \code{data_hist}, which has no default, and the formula of the model, \code{prog_formula}. The default value \code{NULL} for \code{prog_formula} creates a formula that assumes a response variable with the same name as specified in the "main" formula argument and models this by main effects of all columns in \code{data_hist}. The function also has arguments \code{cv_prog_folds} and \code{learners}. \code{cv_prog_folds} specifies the number of cross validation folds during model fitting and testing, while \code{learners} are a list of models to fit for the Discrete Super Learner.

The fitting of the prognostic model to historical data is done by the \newline\code{fit_best_learner()} function, which is also an exported function from \pkg{postcard}. It fits a Discrete Super Learner by taking arguments \code{data_hist}, \code{prog_formula}, \code{cv_prog_folds} and \code{learners} to then use packages from the tidymodels collection \citep{tidymodels2020} to find the model with the lowest test RMSE among specified learners using cross validation.

The object returned by \code{rctglm_with_prognosticscore} inherits the \code{rctglm} class but also has class \code{rctglm_prog}, which has a method \code{prog()} to access the information about the prognostic model. The result of \code{prog()} returns an object of class \code{prognostic_info} with information on \code{formula}, \code{model_fit}, \code{learners}, \code{cv_folds}, and \code{data}.

\subsubsection{Default behavior}

We start by simulating some historical data using the \code{glm_data()} function as before, but now with no treatment variable as the population of historical participants should consist of control participants.

\begin{Schunk}
\begin{Sinput}
R> dat_gaus_hist <- glm_data(
+    Y ~ b0+b1*log(X),
+    X = runif(n, min = 1, max = 50),
+    family = gaussian # Default value
+  )
\end{Sinput}
\end{Schunk}

As mentioned, a call to \code{rctglm_with_prognosticscore()} requires the same arguments as to \code{rctglm()} but with an added specification of the historical data. Thus, a default call which estimates the ATE, adjusting for a prognostic score, is seen below:

\begin{Schunk}
\begin{Sinput}
R> ate_prog <- rctglm_with_prognosticscore(
+    formula = Y ~ A * X,
+    exposure_indicator = A,
+    exposure_prob = 1/2,
+    data = dat_gaus,
+    data_hist = dat_gaus_hist)
\end{Sinput}
\end{Schunk}

\begin{Schunk}
\begin{Sinput}
R> summary(ate_prog)
\end{Sinput}
\end{Schunk}

\begin{Schunk}
\begin{Soutput}
Summary of rctglm fit

Call: rctglm_with_prognosticscore(formula = Y ~ A * X,
exposure_indicator = A,
exposure_prob = 1/2, data = dat_gaus, data_hist = dat_gaus_hist)

Counterfactual control mean (psi_0=E[Y|X, A=0]): 9.958
Counterfactual active mean (psi_1=E[Y|X, A=1]): 12

Estimand function r: psi1 - psi0
Null value (r(psi, psi)): 0

Estimate Std. Error z value Pr(>|z|) 2.5 % 97.5 %
Estimand 2.04203 0.06484 31.49 1.14e-217 1.915 2.17

(Wald-type 95% CI; standard error computed from the full-model
influence-function variance.)
Access the underlying GLM summary via `<summary_object>$glm_summary`.
\end{Soutput}
\end{Schunk}

\subsubsection{Specifying own list of trained learners}

The user can specify a list of any tidymodels \citep{tidymodels2020} as the `learners` argument. Below is an example of fitting the prognostic model as a Discrete Super Learner with the best RMSE among a \textit{random forest} and \textit{linear support vector machines} model.

\begin{Schunk}
\begin{Sinput}
R> learners <- list(
+    rf = list(
+      model = parsnip::set_engine(
+        parsnip::rand_forest(
+          mode = "regression",
+          trees = 500,
+          min_n = parsnip::tune("min_n")
+        ),
+        "ranger"),
+      grid = data.frame(
+        min_n = 1:10
+      )
+    ),
+    svm.linear = list(
+      model = parsnip::set_engine(
+        parsnip::svm_linear(
+          mode = "regression",
+          cost = parsnip::tune("cost"),
+          margin = parsnip::tune("margin")),
+        "LiblineaR"),
+      grid = data.frame(
+        cost = 1:5,
+        margin = seq(0.1, 0.5, 0.1)
+      )
+    )
+  )
R> model_own_learners <- rctglm_with_prognosticscore(
+    formula = Y ~ A * X,
+    exposure_indicator = A,
+    exposure_prob = 1/2,
+    data = dat_gaus,
+    data_hist = dat_gaus_hist,
+    learners = learners)
\end{Sinput}
\end{Schunk}

Recall that the prognostic model can be inspected using the \code{prog} method, which produces an object of class \code{prognostic_info}.

\begin{Schunk}
\begin{Sinput}
R> prog(model_own_learners)
\end{Sinput}
\end{Schunk}

\begin{Schunk}
\begin{Soutput}
Object of class prognostic_info

Formula: Y ~ .
Number of candidate learners: 2
Learner names: rf, svm.linear
CV folds: 5
Historical data: 1000 rows, 2 columns
\end{Soutput}
\end{Schunk}

\section[Prospective power estimation with postcard]{Prospective power estimation with \pkg{postcard}}\label{sec:pros_power_package}

The other main feature of the package is to use historical data to perform prospective power estimation, see Figure~\ref{fig:pkg-overview}. The package contains functionalities for power estimation under the assumption of a correctly specified linear model but here we will focus on the procedure robust to misspecification described in section~\ref{seq:pro-power}. This is implemented in the function \code{power_marginaleffect()}. For linear models the functions \code{variance_ancova()}, \code{power_gs()}, \code{samplesize_gs()} and \code{power_nc()} are available in the package. These also return \code{postcard_power} objects with the same methods as described below. See documentation and vignettes in the package for details on the use of these functions.

The first argument to \code{power_marginaleffect()}, \code{response}, is given as a vector of the response variable in the historical data. \code{predictions} is a vector of predictions of the response obtained from either the GLM model intended for the final analysis of new clinical data, or the fitted prognostic model if the planned final analysis utilises prognostic score adjustment. Both of these arguments have no default value. The same is true for the third and fourth argument, \code{target_effect} and \code{exposure_prob}. \code{target_effect} is the minimum effect size we should be able to detect, $\Psi$, while \code{exposure_prob} is the probability of being in the treatment group, $\pi_1$.

The power functions return objects of class \code{postcard_power}, which hold the estimated power together with the assumptions used to calculate it (sample size, target effect, superiority margin, exposure probability and significance level). The \code{summary} method additionally shows the estimand function, and the \code{plot} method shows the approximate distributions of the estimator under the null and alternative hypotheses, with the area corresponding to the power shaded.

To showcase a simple usage of the function where the planned analysis include the prognostic score adjustment, we create some predictions from the simulated historical data from earlier to then obtain an estimation of the power in a new trial. This is done by first splitting the historical data into a test and training data set for prognostic model building.

\begin{Schunk}
\begin{Sinput}
R> test_gaus <- dat_gaus_hist[1:(nrow(dat_gaus_hist)/10), ]
R> train_gaus <- dat_gaus_hist[
+    (nrow(test_gaus)+1):nrow(dat_gaus_hist),
+    ]
R> lrnr <- fit_best_learner(list(mod = Y ~ X), data = train_gaus)
R> preds <- dplyr::pull(predict(lrnr, new_data = test_gaus))
R> pow_dsl <- power_marginaleffect(
+    response = test_gaus$Y,
+    predictions = preds,
+    target_effect = 1.3,
+    exposure_prob = 1/2
+  )
R> summary(pow_dsl)
\end{Sinput}
\begin{Soutput}
Summary of power approximation (class postcard_power)

Call:  power_marginaleffect(response = test_gaus$Y, predictions = preds, 
    target_effect = 1.3, exposure_prob = 1/2)
Estimated power:      0.8574
Sample size:          100
Target effect:        1.3
Superiority margin:   0
Exposure probability: 0.5
Significance level:   0.05
Estimand function:    psi1 - psi0
\end{Soutput}
\end{Schunk}

\begin{Schunk}
\begin{Sinput}
R> plot(pow_dsl)
\end{Sinput}
\end{Schunk}
\includegraphics{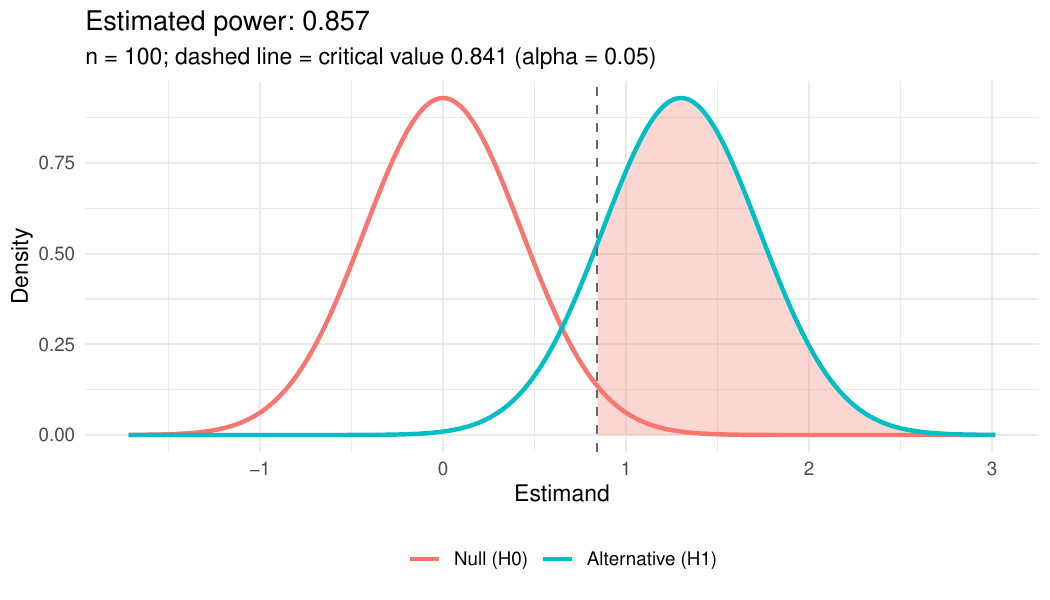}

In the above, power is approximated in the case of using a prognostic model, corresponding to the power of the marginal effect estimated by\newline\code{rctglm_with_prognosticscore}. Using the function to estimate the power without prognostic score adjustment, corresponding to the analysis performed by \code{rctglm}, boils down to finding the predictions from the a GLM rather than from a prognostic model:

\begin{Schunk}
\begin{Sinput}
R> ancova <- glm(Y ~ X, data = train_gaus)
R> preds <- predict(ancova, newdata = test_gaus)
R> power_marginaleffect(
+    response = test_gaus$Y,
+    predictions = preds,
+    target_effect = 1.3,
+    exposure_prob = 1/2
+  )
\end{Sinput}
\begin{Soutput}
Power approximation (class postcard_power)

Call:  power_marginaleffect(response = test_gaus$Y, predictions = preds, 
    target_effect = 1.3, exposure_prob = 1/2)
Estimated power:      0.7962
Sample size:          100
Target effect:        1.3
Superiority margin:   0
Exposure probability: 0.5
Significance level:   0.05
\end{Soutput}
\end{Schunk}

Since \code{response} is the response in historical data of control participants, it corresponds to the potential outcome $Y(0)$. It is used to estimate the variance of the potential outcome under control, $\sigma_0^2$, as well as used together with the predicted outcomes in \code{predictions} to estimate the MSE $\kappa_0^2$. The next two arguments in the function are \code{var1} and \code{kappa1_squared}. These correspond to $\sigma_1^2$ and $\kappa_1^2$, being the variance and MSE, respectively, in the treatment group. As a default, these are estimated by assuming the same value in the control and treatment groups, ie. we set $\hat{\sigma}_1^2=\hat{\sigma}_0^2$ and $\hat{\kappa}_1^2=\hat{\kappa}_0^2$. The user can override this default behavior by specifying \code{var1} or \code{kappa1_squared} as either a \code{numeric} vector of values, or by specifying a function that will perform a transformation of the estimated control group counterpart. 

An argument \code{margin} is used to specify the superiority margin $\Delta$. As a default, no treatment difference is assumed; e.g. a margin of 0 for the ATE estimand and 1 for an RR estimand. The argument \code{alpha} specifies the significance level, using $1-\alpha/2$ as the quantile for the critical value as described in item~\ref{itm:crit_val} in section~\ref{seq:pro-power}.

An example of specifying a superiority margin of 1 in the power estimation for the ATE estimand as well as specifying the variance and MSE of the potential outcomes corresponding to the treatment groups is seen below. We assume that $\sigma_1^2$ is 1.2 times the magnitude of the estimation of $\sigma_0^2$ in the historical data, and we showcase that the MSE $\kappa_1^2$ can be specified as a \code{numeric} value. Both of which in practice might stem from some prior knowledge or belief about the data.

\begin{Schunk}
\begin{Sinput}
R> power_marginaleffect(
+    response = test_gaus$Y,
+    predictions = preds,
+    var1 = function(var0) 1.2 * var0,
+    kappa1_squared = 2,
+    margin = 1,
+    target_effect = 1.3,
+    exposure_prob = 1/2
+  )
\end{Sinput}
\begin{Soutput}
Power approximation (class postcard_power)

Call:  power_marginaleffect(response = test_gaus$Y, predictions = preds, 
    target_effect = 1.3, exposure_prob = 1/2, var1 = function(var0) 1.2 * 
        var0, kappa1_squared = 2, margin = 1)
Estimated power:      0.08981
Sample size:          100
Target effect:        1.3
Superiority margin:   1
Exposure probability: 0.5
Significance level:   0.05
\end{Soutput}
\end{Schunk}

Similar to \code{rctglm()} the \code{power_marginaleffect()} function also has arguments \code{estimand_fun}, \code{estimand_fun_deriv0} and \code{estimand_fun_deriv1}. These have the same behavior for both functions. Additionally, \code{power_marginaleffect()} has an \code{inv_estimand_fun} argument which specifies the inverse estimand functions. If specified, this is used to derive the population mean outcome in the treatment group corresponding to a treatment difference between the groups of \code{target_effect}, ie. determine $\Psi_1$ such that $r(\Psi_1, \Psi_0)=$\code{target_effect}. If \code{inv_estimand_fun} is not specified, a numeric derivation of $\Psi_1$ is performed. The \code{tolerance} argument controls the sensitivity when checking whether the numeric derivation of $\Psi_1$ produced good results and is thus only relevant when \code{inv_estimand_fun} is \code{NULL}.

\subsection{Creating a plot of prospective power curves}

To visualise how the estimated power behaves as a function of the total sample size compared between models, functions \code{repeat_power_marginaleffect()} and \code{repeat_power_linear()} are available, which both produce S3 class objects with associated plot methods. We will here describe and show an example for \code{repeat_power_marginaleffect()}. While the arguments are slightly different, the methodology of \code{repeat_power_linear()} is the same.

The function has two arguments with non-default values. Like in\newline\code{power_marginaleffect()}, we need to specify our \code{target_effect} and an\newline\code{exposure_prob}. Arguments \code{model_list} and \code{test_data_fun} do have default values just to enable an easy way for the user to inspect the output of the function, but these are arguments the user will typically want to specify. These arguments give a list of fitted models, which are used to create predictions on the test data created for each sample size, which are then passed to \code{power_marginaleffect()}. In this article, we create data sets and fit models that correspond to the default behavior to be explicit about how the user could define and pass these arguments.

The function also has arguments \code{ns}, \code{desired_power} and \code{n_iter}, which are a vector of sample sizes to estimate the power for, the power that is trying to be achieved, and a number of iterations to average the results over, respectively.

We start by creating some data to fit the models we want to test the power for, and defining a function with a single argument to generate test data for each sample size. We generate data with a non-linear effect on the covariate \code{X}, and with an interaction effect between \code{X1} and \code{X2}.

\begin{Schunk}
\begin{Sinput}
R> n_train <- 2*1e3
R> train_gaus <- glm_data(
+    Y ~ b0+b1*log(X),
+    X = runif(n_train, min = 1, max = 50),
+    family = gaussian # Default value
+  )
R> test_gaus_fun <- function(n) {
+    glm_data(
+    Y ~ b0+b1*log(X),
+    X = runif(n, min = 1, max = 50),
+    family = gaussian # Default value
+  )
+  }
\end{Sinput}
\end{Schunk}

We then create a list of models that we want to estimate the power for. Here, we choose an ANCOVA adjusts linearly for all the covariates in the data generating process as well as an ANCOVA that utilises prognostic score adjustment, with prognostic scores from a prognostic model fitted as a discrete super learner with the function \code{fit_best_learner}. Note that we use the conservative power estimation approach described in \citep{HojbjerreFrandsen2025} for the ANCOVA model that utilises prognostic score adjustment by solely using the prognostic scores as predictions rather than the ANCOVA model that adjust for the prognostic score as well as other covariates.

\begin{Schunk}
\begin{Sinput}
R> ancova_prog_list <- list(
+    ANCOVA = glm(
+      Y ~ X,
+      data = train_gaus),
+    "ANCOVA with prognostic score" = fit_best_learner(
+      list(mod = Y ~ X),
+      data = train_gaus)
+  )
\end{Sinput}
\end{Schunk}

We run the power estimation, specifying arguments that are passed to \code{power_marginaleffect()}.

\begin{Schunk}
\begin{Sinput}
R> rpm <- repeat_power_marginaleffect(
+   model_list = ancova_prog_list,
+   test_data_fun = test_gaus_fun,
+   ns = seq(10, 300, 1), n_iter = 50,
+   var1 = function(var0) 1.1 * var0,
+   kappa1_squared = function(kap0) 1.1 * kap0,
+   target_effect = 0.8,
+   exposure_prob = 1/2,
+   margin = -0.2
+  )
\end{Sinput}
\end{Schunk}

The result is an object of class \code{postcard_rpm} containing information on the estimated power for each sample size. Its \code{summary} method shows the assumptions used for the power estimation, including the estimand function, along with information on for which sample size the desired power was reached, what the power at the biggest tested sample size is, and the maximal power achieved among all cases.

\begin{Schunk}
\begin{Sinput}
R> summary(rpm)
\end{Sinput}
\begin{Soutput}
Summary of power curves (class postcard_rpm)

Power approximation:  power_marginaleffect
Sample sizes:         10 to 300
Desired power:        0.9
Target effect:        0.8
Superiority margin:   -0.2
Exposure probability: 0.5
Significance level:   0.05

                        model n_achieve_power power_at_max_n
                       ANCOVA             244         0.9485
 ANCOVA with prognostic score             201         0.9746
 max_power
    0.9518
    0.9774
\end{Soutput}
\end{Schunk}

Its \code{plot} method shows the approximate distributions of the estimator under the null hypothesis (centered at the \code{margin}) and the alternative hypothesis (centered at the \code{target_effect}), with the area corresponding to the power shaded.

\begin{Schunk}
\begin{Sinput}
R> plot(rpm)
\end{Sinput}
\end{Schunk}
\includegraphics{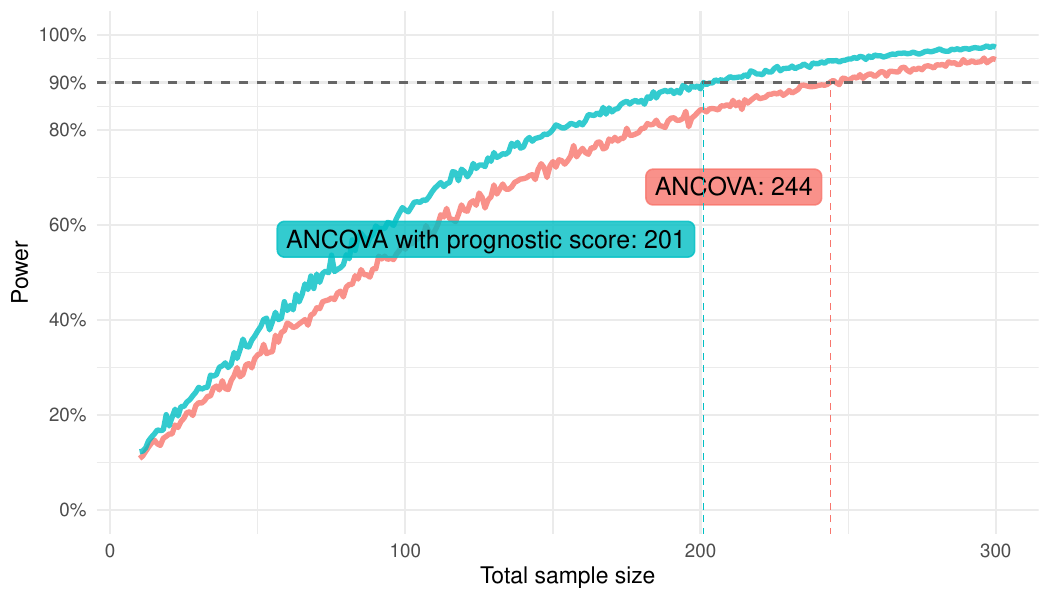}

In the above figure we see that the ANCOVA model with prognostic score adjustment achieves the desired power of 90\% with a sample size of 201, while the "regular" ANCOVA reaches the desired power at a sample size of 244. 

\section{Conclusion}

\pkg{postcard} aims to make estimation of marginal effects in RCTs with plug-in GLM and prognostic score adjustment straightforward and accessible in \proglang{R}. In addition to the using plug-in estimation, IFs are used to obtain variance estimates. The package implements functions \code{rctglm} and \code{rctglm_with_prognosticscore} with interfaces similar to \code{glm} allow users to familiarize themselves with the package quickly. These functions return objects that contain the marginal effect estimand, marginal effect estimates, their variances, the underlying \code{glm} fit, and for \code{rctglm_with_prognosticscore} information about the prognostic model fitted as a Discrete Super Learner. 

The package also provides functions for statistical power approximation, including methods for linear models but more importantly methods that remain valid even when model assumptions are misspecified. This enables users to approximate the power of any marginal effect for any model. This thus includes the approximation of power resulting from an analysis using prognostic score adjustment.

A function for generating data following a GLM is also included in the package, allowing users to perform their own exploratory analyses and simulation studies.

By combining estimation, prognostic adjustment, and power approximation in a single, user-friendly toolkit, postcard supports both the analysis and planning phases of clinical research. With increasing emphasis on precision, reproducibility, and transparent reporting in RCTs, postcard provides a practical and efficient way to implement state-of-the-art covariate adjustment methods in applied settings.



\section*{Computational details}
The results in this paper were obtained using R 4.6.1 and development version 1.1.0.9000 (tag \href{https://github.com/NovoNordisk-OpenSource/postcard/releases/tag/v1.1.0.9000-paper}{v1.1.0.9000-paper}) of the postcard package. R itself
and all packages used are available from the Comprehensive R Archive Network (CRAN) at
https://CRAN.R-project.org/.

\section*{Acknowledgments}
This research was supported by Innovation Fund Denmark (grant number 2052-00044B). Furthermore, thanks to Claus Dethlefsen for useful feedback on the article.

\bibliographystyle{plainnat}
\bibliography{refs}

\end{document}